# Photoinduced ultrafast transition of the local correlated structure in chalcogenide phase-change materials

Yingpeng Qi[1,3*], Nianke Chen[2], Thomas Vasileiadis[1], Daniela Zahn[1], Hélène Seiler[1], Xianbin Li[2*], Ralph Ernstorfer[1*]

[1]Fritz-Haber-Institut der Max-Planck-Gesellschaft, Faradayweg 4-6, Berlin 14195, Germany

[2]State Key Laboratory of Integrated Optoelectronics, College of Electronic Science and Engineering, Jilin University, 2699 Qianjin Street, Changchun 130012, China

[3]Center for Ultrafast Science and Technology, School of Physics and Astronomy, Shanghai Jiao Tong University, 200240 Shanghai, China

*Correspondence to qiyp@sjtu.edu.cn, lixianbin@jlu.edu.cn and ernstorfer@fhi-berlin.mpg.de

## Abstract

Revealing the bonding and time-evolving atomic dynamics in functional materials with complex lattice structures can update the fundamental knowledge on rich physics therein, and also help to manipulate the material properties as desired. As the most prototypical chalcogenide phase change material, $Ge_2Sb_2Te_5$ has been widely used in optical data storage and non-volatile electric memory due to the fast switching speed and the low energy consumption. However, the basic understanding of the structural dynamics on the atomic scale is still not clear. Using femtosecond electron diffraction, structure factor calculation and TDDFT-MD simulation, we reveal the photoinduced ultrafast transition of the local correlated structure in the averaged rock-salt phase of $Ge_2Sb_2Te_5$. The randomly oriented Peierls distortion among unit cells in the averaged rock-salt phase of $Ge_2Sb_2Te_5$ is termed as local correlated structures. The ultrafast suppression of the local Peierls distortions in individual unit cell gives rise to a local structure change from the rhombohedral to the cubic geometry within ~ 0.3 ps. In addition, the impact of the carrier relaxation and the large amount of vacancies to the ultrafast structural response is quantified and discussed. Our work provides new microscopic insights into contributions of the local correlated structure to the transient structural and optical responses in phase change materials. Moreover, we stress the significance of femtosecond electron diffraction in revealing the local correlated structure in the subunit cell and the link between the local correlated structure and physical properties in functional materials with complex microstructures.

Due to the growth of the global amount of data, the demand for data storage and processing is increasing exponentially [1]. Chalcogenide phase change materials (PCMs) have been singled out as one of the best classes of prospective materials for all-photonic storage/memory [2] and electronic phase-change memory [1]. A laser or electrical pulse stimulates nonvolatile switching between a crystalline and an amorphous phase with atypically large differences in the optical-reflectivity and the electrical-resistivity. During such a structural phase switching, the crystallization process is the time-limiting step, while the amorphization process is the energy-intensive step. To optimize these two steps, one hot topic is to modulate the local structural geometry in the atomic level in the most prototypical phase-change material $Ge_2Sb_2Te_5$ (GST-225) and the similar alloys along the tie-line of $GeTe-Sb_2Te_3$ [3-9]. The crystallization speed has been improved from tens of nanoseconds to sub-nanosecond by introducing prestructural ordering [4] and crystal precursors [5]. On the other hand, the energy consumption for the amorphization can be reduced by introducing the premelting disordering [6] and controlling the local atomic switching [7]. The recent theoretical studies [8, 9] propose that an ultrafast electronic excitation can reduce the energy consumption by introducing a direct solid-solid amorphization bypassing the molten state in GST-225. Overall, engineering the local structures in GST-225 play a key role in improving the performance of the structural switching.

The crystalline GST-225 is an averaged rock-salt phase, accompanied by local distortions and huge vacancy concentrations, as shown in Fig. 1a. The anion sub-lattice is fully occupied by Te atoms, whereas the cation sites are populated by Ge (40%), Sb (40%) and vacancies (20%) randomly [3]. The local structure in individual unit cell of GST-225 is rhombohedrally distorted with random orientation to eight equivalent distortions along the <111> directions [9, 10, 11]. Such local distortions with shorter and longer Ge(Sb)-Te bonds is termed as local Peierls distortions as shown in Fig. 1a (bottom). From a chemical perspective, the half-filled p-band of Ge (Sb/Te) forms two bonds to the left and right atoms, and this bonding structure is termed resonant bond [12], metavalent bond [13], or multicenter hyperbonding [14]. The randomly oriented Peierls distortions in GST-225 are a typical characteristic of local correlated structures in crystalline functional materials [15-20]. What is common to material systems with local correlated structures is that there exists a distinction between the local symmetry and the average symmetry imposed by the crystal lattice [15]. Photoexcitation of GST-225 triggers a drastic optical contrast within 100-200 femtoseconds, which has been attributed to the depletion of electrons from the metavalent bonds [21, 22]. Besides such an electronic structure modulation, breaking the bond alignment [23] and the medium range order of the lattice structure [24] may also induce a significant change of the optical matrix elements. However, the precise knowledge on the ultrafast structural response is still not clear. The atomic motions after femtosecond laser excitation have been studied extensively by time-resolved molecular dynamics simulation, ultrafast spectroscopy and ultrafast electron/X-ray diffraction. Nevertheless, the developed structural response models, including the phonon-driven symmetry change [25, 26], the rattling motion [23], the selective bond breaking [8, 9] and the simple thermal response [21, 22], do not provide a consistent picture. The local Peierls distortions in the rock-salt phase of GST-225, which may associate with the highly damped Raman-inactive phonon mode in Ref. 22, is always neglected in previous studies.

Ultrafast electron/X-ray probes enable a direct access to transient atomic and electronic motions in a broad range of fundamental physical processes after femtosecond laser excitation [27-35]. Here we report the photo-induced local structural dynamics in phase change material GST-225 revealed with a combination of femtosecond electron diffraction, structure factor calculation and time-dependent density-functional theory molecular dynamic simulations (TDDFT-MD). The high temporal resolution of ~ 150 fs in our experimental system [36, 37], which is comparable to the highest vibration period of phonons in GST-225 [22], enables a direct access to the ultrafast structural response induced by electronic excitation. A consistent physical scenario towards an ultrafast transition of the local correlated structure is unveiled.

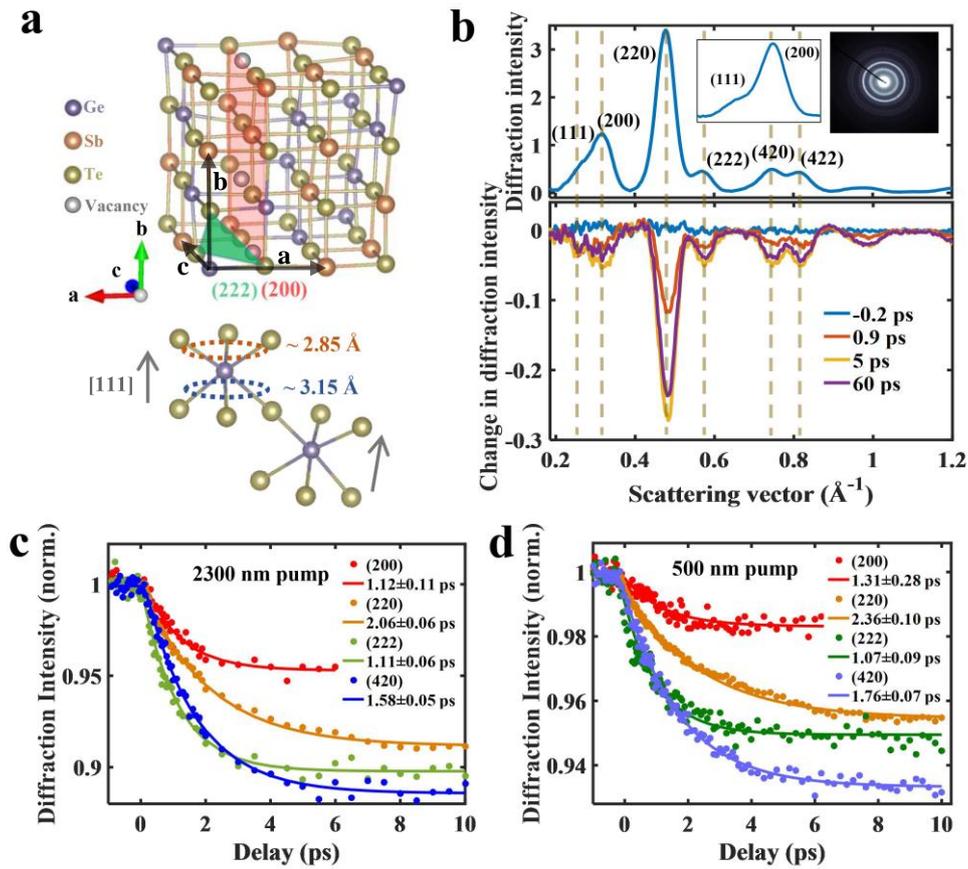

Fig. 1. The crystalline structure and the transient structural dynamics of GST-225. (a) (top) A supercell structure of the rock-salt phase GST-225. The length of three axes of single unit cell is 6.01 Å. (bottom) A schematic illustration of randomly oriented Peierls distortions with longer (~ 3.15 Å) and shorter (~ 2.85 Å) Ge(Sb)-Te bonds in the subunit cell. The gray arrows indicate the orientation of the adjacent local rhombohedral structures. (b) (top) The transmission diffraction pattern and the radially averaged intensity distribution (background subtracted) at a negative delay. The inset displays a better separation of (111) and (200) reflections by controlling the current of the magnetic lens positioned after the sample. (bottom) The change in diffraction intensity from -0.2 to 60 ps, by subtracting the intensity distribution at negative delay with 2300 nm and 4.6 mJ/cm$^2$ laser pump. (c) (d) The temporal evolution of the normalized Bragg reflection intensities with 2300 nm (4.6 mJ/cm$^2$) and 500 nm (0.61 mJ/cm$^2$) laser pump. The solid lines are the fits with a monoexponential function.

In the experiment, we use a femtosecond laser pulse to excite the GST-225 nanofilm with a thickness of 15 nm (see details in Supplementary Material [38]). A femtosecond electron pulse diffracts off the transient lattice structure at varying time delays. The transmission diffraction pattern and the radially averaged intensity distribution of the rock-salt phase GST-225 are shown in Fig. 1b (top). The change in diffraction intensity in Fig. 1b (bottom) shows a transient decrease of the Bragg reflection intensities after femtosecond laser excitation. To analyze the detailed temporal evolution, we fit an empirical back-ground function (exponential plus second-order polynomial) and pseudo-Voigt line profiles to the peaks in the radial averages at each pump-probe delay [55].

Since the band gap of GST-225 is 0.4 ~ 0.5 eV [56], we choose to pump with two different wavelengths, i.e. 2300 nm (0.54 eV) and 500 nm (2.48 eV), in order to investigate the possible impact of the carrier relaxation to the structural response [57]. The relative intensity changes of several Bragg reflections with 2300 nm and 500 nm optical pump are depicted in Fig. 1c and Fig. 1d. We fit the temporal evolution of the intensity with a monoexponential function, convolved with a Gaussian function of 150 fs FWHM to account for the instrument response function. For each reflection, the time constant is nearly the same for 500 nm and 2300 nm laser excitation. With increasing pump fluence, the temporal evolution of the intensities and the corresponding time constants are summarized in Fig. S1-S2 and Table 1-2 in Supplementary Material. In these measurements, the density of the excited electrons, i.e. $(0.69-2.43)\times10^{15}$ cm$^{-2}$, with 2300 nm laser pump (~2-7 mJ/cm$^2$) is comparable with the density of $(0.59-1.57)\times10^{15}$ cm$^{-2}$ for 500 nm laser excitation (~0.3-0.8 mJ/cm$^2$) (see density calculation in Supplementary Material). The similar structural response with 500 nm and 2300 nm laser excitation is obtained. Whether the carrier relaxation will impact the amorphization process at a higher pump fluence [57] needs to be studied further, but at the low to medium pump fluence, we do not observe distinct structural responses with low to high photon energy excitation.

The local Peierls distortion is randomly oriented along the <111> directions in the averaged rock-salt phase of GST, therefore, we focus on the intensity change of the (111) reflection as shown in Fig. 2. To avoid the possible impact of the heat accumulation [58], experiments at room temperature (Fig. 2a) and at 112 K (Fig. 2b) are performed. The temporal evolution of the intensities in Fig. 2a and 2b clearly indicate an ultrafast intensity decay of the (111) reflection. See detail on the data processing in Fig. S3-S6. The time constant of 0.1 - 0.3 ps for the (111) reflection is significantly smaller than other Bragg reflections in Fig. 1, which may relate to the local Peierls distortion. In conventional crystalline material, the Peierls distortion is generally induced by electronic instability, and a femtosecond laser excitation gives rise to a suppression of such structural distortion by coherent phonons [25, 27, 29]. Regarding GST-225, the Peierls distortions are local and randomly oriented along the <111> directions, therefore, the suppression of adjacent local Peierls distortions across unit cells should be incoherent as illustrated in Fig. 2c. In particular, the photoexcitation flattens the potential energy surface along the <111> directions and enables the opposite movement of the Ge(Sb) and Te atom, giving rise to the local structural transition from the rhombohedral to the cubic geometry. With different pump fluences, the time constant for the intensity decay of the (111) reflection is 0.1 - 0.3 ps as shown in Fig. 2d, smaller than the period of the $A_{1g}$ phonon mode in GeTe [25]. A quasi-linear dependence of the amplitude of the (111) intensity change on the pump fluence is shown in Fig. S7.

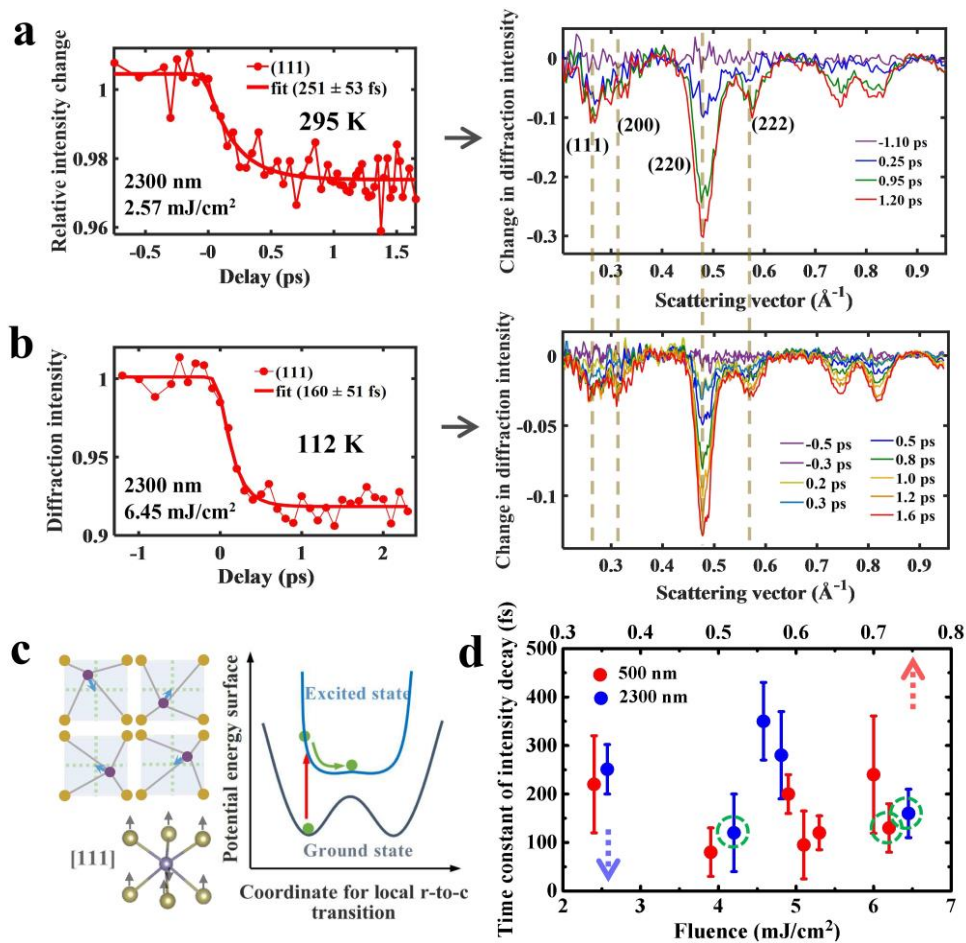

Fig. 2. (a) (b) with 2300 nm laser pump, (left) Normalized intensity change of the (111) reflection as a function of the time delay at room temperature and 112 K respectively. The solid line is the fit with a monoexponential function; (right) Overall change in diffraction intensity from negative to positive delays. (c) The 2D square nets indicates the suppression of the local distortions with random orientation among unit cells. The Ge/Sb atoms (purple balls) move towards the high symsmetry position (the corresponding movement of Te atom is not shown for simplification). In the subunit cell, the sketch of the atomic displacement along the [111] direction and the corresponding potential energy surface change are shown. (d) Time constants of the intensity decay of the (111) reflection with different pump fluences for 500 nm and 2300 nm laser excitation at room temperature and 112 K. The error bars correspond to 68.3% confidence intervals of the fit. The measurements carried out at 112 K are marked with green balls.

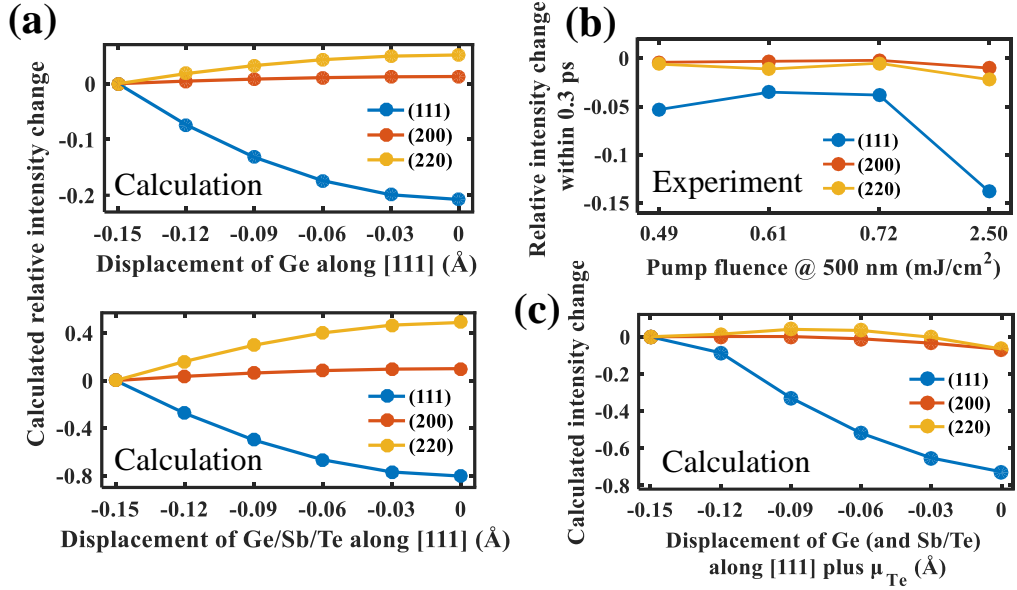

Fig. 3. Calculated and experimental anisotropic intensity change associated with the suppression of the local rhombohedral structure. (a) By structure factor calculation, the relative intensity change of Bragg reflections ($|F|^2$) with (top) only the offset displacement of Ge atom and (bottom) the opposite displacement of Ge/Sb and Te atom along the [111] direction. The zero displacement position indicates the site in an ideal rock-salt phase. The offset displacement of atoms along the [111] direction represents the local rhombohedral structure in individual unit cell. (b) The measured anisotropic intensity change within 0.3 ps, i.e. the larger intensity change of the (111) reflection than that of (200) and (220) reflection. (c) Calculated intensity change with the displacement of Ge from -0.15 Å to 0 Å, Sb from -0.08 Å to 0 Å, Te from 0.08 Å to 0 Å, and the vibrations of Te atoms ($\mu_{Te}$ ~ 0 to 0.2 Å).

To quantify the suppression of the local rhombohedral structure, we perform the structure factor calculation. Note that for crystalline materials with local correlated structures, such as the randomly oriented Peierls distortions in adjacent unit cell in GST-225, the diffraction intensity is calculated by a summation over all unit cells included. If the atomic motions associated with such local distortions are uncorrelated across unit cells, the structure factor calculation of a single unit cell can be used to evaluate the diffraction intensity change of the overall lattice. See details about the structure factor calculation in Supplementary Material. Fig. 3a shows the calculation results on two displacement models, indicating the transition from the rhombohedral to the cubic configuration (-0.15 Å to 0 Å). In the first model, only the offset displacement for the Ge atom along the [111] direction is considered because the lighter Ge atom is expected to move more easily than the heavier Te and Sb atoms, while in the second model, the opposite displacement of Ge/Sb and Te atom along the [111] direction is used. The same anisotropic intensity change for the two models is observed, i.e. the intensity of the (111) reflection is decreased significantly while the intensity of the (200) and (220) reflection is remained or enhanced. In Fig. 3b, we extract the experimental intensity change within 0.3 ps, where the intensity decay of the (111) reflection is almost done. As seen, the intensity of the (111) reflection is decreased remarkably while the intensity for the (200) and (220) reflection remains, which is qualitatively in agreement with the structure factor calculation in Fig. 3a.

For a more precise structure factor calculation, two points need to be considered. First, in the averaged rock-salt phase of GST-225, the length of the shorter Ge-Te bond in Ge-centered local rhombohedral structure is ~ 2.84 Å (~ 2.82 Å), while the shorter Sb-Te bond in Sb-centered local rhombohedral structure is ~ 2.91 Å (~ 2.88 Å) at 100 k (300 k) [9]. Second, with electronic excitation, the strong local force would drive large vibrations of Te atoms around vacuum sites within 0.5 ps [59]. With the combination of the bond length discrepancy and the local vibrations of Te atoms (see details in Supplementary Material), the calculated intensity change in Fig. 3c agrees better with the experimental results in Fig. 3b.

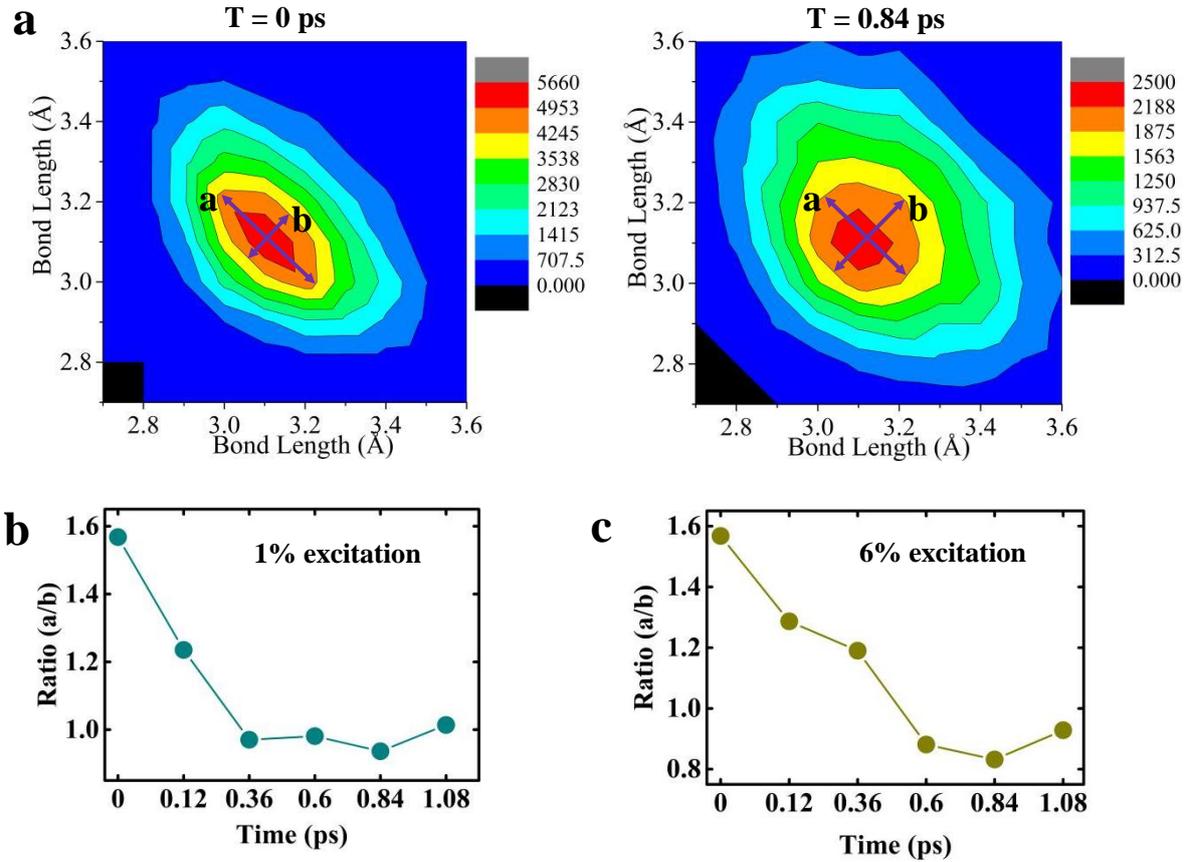

Fig. 4. Simulation results of bond length in linear triatomic bonding geometry in the supercell of GST-225 before and after the electronic excitation. (a) (left) The bond length distribution within the supercell before laser excitation. The tilted distribution indicates a shorter and a longer bond in a linear triatomic bonding geometry. (right) The bond length distribution at 0.84 ps (integration from 0.72 ps to 0.96 ps) after 6% electronic excitation. (b) (c) The temporal evolution of the ratio between the length of line a and line b (indicated in (a)) with 1% and 6% electronic excitation respectively. The decrease of the the ratio, i.e. the ellipticity, indicates the suppression of the bond length discrepancy in a linear triatomic bonding geometry.

We further confirm the photo-induced suppression of local Peierls distortions by performing TDDFT-MD simulations in the averaged rock-salt phase of GST-225. See detailes about the simulation in Supplementary Material. The photoexcitation effect is simulated by moving a certain percent of electrons from the valence band maximum states to the conduction band minimum states, similar to the method in Ref. 25. In order to characterize the bond length discrepancy in the supercell, the bond pairs in each linear triatomic bonding geometry, such as the Te-Ge-Te and the Te-Sb-Te, are tracked during the temporal evolution. The contour map in Fig. 4a displays the distribution of the collected bond pairs. At the ground state (T=0 ps), the elliptical distribution of the contour map in Fig. 4a (left) indicates the shorter and the longer bond in most bond pairs, an intrinsic character of the local Peierls distortions. After electronic excitation, as shown in Fig. 4a (right), the ellipticity of the contour map is reduced, i.e. the bond length discrepancy in bond pairs is reduced. We quantify the change of the bond length discrepancy by calculating the ellipticity of the contour map, i.e. the ratio between the length of line a and line b for the brown zone. The maximum electronic excitation in our experiment is ~ 1% (see the calculation in Supplementary Material), therefore we perform 1% and 6% electronic excitation in the simulation. The temporal evolutions of the ellipticity are displayed in Fig. 4b and Fig. 4c (each data point is the statistic over ± 120 fs). The ellipticity has decreased apparently from 1.6 to around 1, indicating the suppression of the local Peierls distortions. Moreover, the change of the ellipticity completes within 0.3~0.6 ps, which agrees with the timescale of the ultrafast intensity decay of the (111) reflection in Fig. 2. Therefore, the simulation result coincides with the physical model of the photo-induced suppression of local Peierls distortions concluded from the experimental results. The time-resolved bond length distribution by counting the Te-Ge-Te and the Te-Sb-Te geometry is shown in Fig. S9.

Using femtosecond electron diffraction, structure factor calculation and TDDFT-MD simulations, we reveal the ultrafast suppression of the local Peierls distortions in the averaged rock-salt phase of GST-225. Distinct from the model of the rattling motion [23], the selective bond breaking [8, 9] and the simple thermal response [21, 22], a local structural transition from the rhombohedral to the cubic geometry within ~ 0.3 ps is demonstrated. Since the Peierls distortion is randomly oriented along the <111> directions, the atomic displacements associated with the suppression of Peierls distortions are coherent within single unit cell while they are incoherent across unit cells. Therefore, such local structural transition is intrinsically different from the conventional long-range symmetry change, such as the rhombohedral-to-cubic transition in GeTe driven by the coherent $A_{1g}$ phonon mode [25]. The highly damped Raman-inactive phonon mode observed in GST-225 in Ref. 22 should be attributed to the ultrafast suppression of the local Peierls distortions instead of the vacancy sites. The timescale of the ultrafast local structural transition is comparable to that of the ultrafast dielectric function change [21, 22], therefore the local structural transition is expected to contribute to the dielectric function change (see discussion in Supplementary Material).

A direct solid-solid amorphization bypassing the molten state has long been pursued in GST-225 [6-9, 57, 59, 60]. In contrast to the enhancement of the Peierls distortions with rising temperature [9], the identified ultrafast suppression of the local Peierls distortions here is photo-induced. The similar ultrafast structural responses with 2300 nm and 500 nm laser pump indicates no significant impact of the carrier relaxation to the structural dynamics [57]. We expect upon increasing the pump fluence, the suppression of the Peierls distortions will be the intrinsic process driving the system towards the amorphization phase, which is distinct from the model of the selective breaking of the longer bonds in the local distortion [9, 61]. See more discussion on the possible direct amorphization based on the ultrafast local structural transition in Supplementary Material.

The local correlated structure (which is also termed as correlated disorder in Ref. 15), such as the randomly oriented Peierls distortions in GST-225 and $BaTiO_3$ (see Ref. 15), is an intrinsic structural character in many functional crystalline materials and is important for the particular function of interest [15, 16]. With conventional crystallography technique (such as static X-ray diffraction) determining the space group, only the long-range order is clearly detected and the local correlated structure is hidden. How to characterize the local correlated structure and establish the link between the physical properties and the disorder is challenging. Intrinsically, such local correlated structures by electronic instability is expected to be released in an ultrafast way after femtosecond laser excitation. The corresponding ultrafast atomic motions can be isolated by monitoring the ultrafast intensity changes of diffraction peaks (we present in this work and in Ref. 62), the diffuse scattering [63] and the pair distribution function. Then the local correlated structures can be visualized and identified directly with time-resolved diffraction method, such as ultrafast electron diffraction. We anticipate that the comprehensive analysis of the ultrafast structural responses in our work will help to reveal the local correlated structure [15-20] in crystalline functional materials, and deepen the understanding of the local structural dynamics in halide perovskites [64, 65] and order-order/disorder structural transitions in thermoelectric materials [66-69] and ferroelectric-paraelectric transitions [70, 71].

## Data and materials availability

All data needed to evaluate the conclusions in the paper are present in the paper and the Supplementary Materials. Materials related to this paper may be requested from the corresponding author Yingpeng Qi or Ralph Ernstorfer.

1. W. Zhang, R. Mazzarello, M. Wuttig and E. Ma, Designing crystallization in phase-change materials for universal memory and neuro-inspired computing, *Nat. Rev. Mater*. **4**, 150 (2019).
2. M. Wuttig, H. Bhaskaran and T. Taubner, Phase-change materials for non-volatile photonic applications, *Nat. Photon*. **4**, 83 (2010).

3. M. Wuttig and N. Yamada, Phase-change materials for rewriteable data storage. *Nat. Mater* **6**, 824 (2007).

4. D. Loke, T. H. Lee, W. J. Wang, L. P. Shi, R. Zhao, Y. C. Yeo, T. C. Chong, S. R. Elliott, Breaking the Speed Limits of Phase-Change Memory, *Science* **336**, 1566 (2012).

5. F. Rao et al., Reducing the stochasticity of crystal nucleation to enable subnanosecond memory writing, *Science* **358**, 1423 (2017).

6. Desmond Loke, Jonathan M. Skelton, Wei-Jie Wang, Tae-Hoon Lee, Rong Zhao, Tow-Chong Chong, and Stephen R. Elliott, Ultrafast phase-change logic device driven by melting processes, *Proc. Natl. Acad. Sci.* **111**, 13272 (2014).

7. R. E. Simpson, P. Fons, A. V. Kolobov, T. Fukaya, M. Krbal, T. Yagi and J. Tominaga, Interfacial phase-change memory. *Nat. Nanotech.* **6**, 501 (2011).

8. X. B. Li, X. Q. Liu, X. Liu, D. Han, Z. Zhang, X. D. Han, Hong-Bo Sun, and S. B. Zhang, Role of Electronic Excitation in the Amorphization of Ge-Sb-Te Alloys. *Phys. Rev. Lett.* **107**, 015501 (2011).

9. A. V. Kolobov, M. Krbal, P. Fons, J. Tominaga, and T. Uruga, Distortion-triggered loss of long-range order in solids with bonding energy hierarchy. *Nat. Chem.* **3,** 311–316 (2011).

10. A. V. Kolobov, P. Fons, A. I. Frenkel, A. L. Ankudinov, J. Tominaga, and T. Uruga, Understanding the phase-change mechanism of rewritable optical media, *Nat. Mater.* **3**, 703 (2004).

11. M. Wuttig, D. Lüsebrink, D. Wamwangi, W. Wełnic, M. Gilleßen, and R. Dronskowski, The role of vacancies and local distortions in the design of new phase-change materials. *Nat. Mater.* 6, 122–128 (2007).

12. K. Shportko, S. Kremers, M. Woda, D. Lencer, J. Robertson and M. Wuttig, Resonant bonding in crystalline phase-change materials, *Nat. Mater.* **7**, 653 (2008).

13. M. Wuttig, V. L. Deringer, X. Gonze, C. Bichara, J. Raty, Incipient Metals: Functional Materials with a Unique Bonding Mechanism, *Adv. Mater.* **30**, 1803777 (2018).

14. T. H. Lee, S. R. Elliott, Chemical Bonding in Chalcogenides: The Concept of Multicenter Hyperbonding, *Adv. Mater.* **32**, 2000340 (2020).

15. A. Simonov and A. L. Goodwin, Designing disorder into crystalline materials, *Nat. Rev. Chem.* **4**, 657 (2020).

16. G. Jeffrey Snyder and Eric S. Toberer, Complex thermoelectric materials, *Nat. Mater.* **7**, 105 (2008).

17. A. J. Sievers, S. Takeno, Intrinsic Localized Modes in Anharmonic Crystals, *Phys. Rev. Lett.* **61**, 970 (1988).

18. D. K. Campbell, S. Flach, Y. S. Kivshar, Localizing energy through nonlinearity and discreteness. *Phys. Today* **57**, 43 (2004).

19. M. E. Manley et al. Intrinsic anharmonic localization in thermoelectric PbSe. *Nat. Commun.* **10**, 1928 (2019).

20. M. E. Manley, A. J. Sievers, J. W. Lynn, S. A. Kiselev, N. I. Agladze, Y. Chen, A. Llobet, and A. Alatas, Intrinsic localized modes observed in the high-temperature vibrational spectrum of NaI, *Phys. Rev. B* **79**, 134304 (2009).

21. L. Waldecker, T. A. Miller, M. Rudé, R. Bertoni, J. Osmond, V. Pruneri, R. E. Simpson, R. Ernstorfer, and S. Wall, Time-domain separation of optical properties from structural transitions in resonantly bonded materials. *Nat. Mater.* **14,** 991–996 (2015).

22. T. A. Miller, M. Rudé, V. Pruneri, and S. Wall, Ultrafast optical response of the amorphous and crystalline states of the phase change material $Ge_2Sb_2Te_5$. *Phys. Rev. B* **94,** 024301 (2016).

23. E. Matsubara *et al.,* Initial Atomic Motion Immediately Following Femtosecond-Laser Excitation in Phase-Change Materials. *Phys. Rev. Lett.* **117***,* 135501 (2016).

24. B. Huang and J. Robertson, Bonding origin of optical contrast in phase-change memory materials, *Phys. Rev. B* **81**, 081204(R) (2010).

25. N. K. Chen, X. B. Li, J. Bang, X. P. Wang, D. Han, D. West, S. B. Zhang, and H. B. Sun, Directional Forces by Momentumless Excitation and Order-to-Order Transition in Peierls-Distorted Solids: The Case of GeTe. *Phys. Rev. Lett.* **120**, 185701 (2018).

26. J. B. Hu, G. M. Vanacore, Z. Yang, X. S. Miao, and A. H. Zewail, Transient Structures and Possible Limits of Data Recording in Phase-Change Materials, *ACS Nano* **97**, 6728-6737 (2015).

27. D. M. Fritz et al. Ultrafast bond softening in Bismuth: mapping a solid's interatomic potential with X-rays, *Science* **315**, 633 (2007).

28. M. Z. Mo et al. Heterogeneous to homogeneous melting transition visualized with ultrafast electron diffraction, *Science* **360**, 1451 (2018).

29. K. Sokolowski-Tinten, C. Blome, J. Blums, A. Cavalleri, C. Dietrich, A. Tarasevitch, I. Uschmann, E. Förster, M. Kammler, M. Horn-von-Hoegen, and D. Von der Linde, Femtosecond x-ray measurement of coherent lattice vibrations near the lindemann stability limit, *Nature* **422**, 287 (2003).

30. R. J. Dwayne Miller, Femtosecond Crystallography with Ultrabright Electrons and X-rays: Capturing Chemistry in Action, *Science* **343**, 1108 (2014).

31. Jie Yang et al. Imaging $CF_3I$ conical intersection and photodissociation dynamics with

ultrafast electron diffraction, *Science* **361**, 64 (2018).

32. M. Eichberger, H. Schäfer, M. Krumova, M. Beyer, J. Demsar, H. Berger, G. Moriena, G. Sciaini, and R. J. D. Miller, Snapshots of cooperative atomic motions in the optical suppression of charge density waves, *Nature* **468**, 799 (2010).

33. S. Gerber et al. Femtosecond electron-phonon lock-in by photoemission and x-ray free-electron laser, *Science* **357**, 71 (2018).

34. Anshul Kogar et al. Light-induced charge density wave in LaTe$_3$, *Nat. Phys.* **16**, 159 (2020).

35. V. R. Morrison et al. A photoinduced metal-like phase of monoclinic VO$_2$ revealed by ultrafast electron diffraction, *Science* **346**, 445 (2014).

36. L. Waldecker, R. Bertoni, and R. Ernstorfer, Compact femtosecond electron diffractometer with 100 keV electron bunches approaching the single-electron pulse duration limit. *J. Appl. Phys.* **117**, 044903 (2015).

37. D. Zahn, P. Hildebrandt, T. Vasileiadis, Y. W. Windsor, Y. Qi, H. Seiler, and R. Ernstorfer, Anisotropic Nonequilibrium Lattice Dynamics of Black Phosphorus. *Nano Lett.* **20**, 3728 (2020).

38. See Supplemental Material for details of experiments and sample preparation, the method for first-principles calculations, the calculation of the density of excited electrons, the experimental raw data and the data processing, the structure factor calculation and the contribution of the local structural transition to the dielectric function change and the direct amorphization, which includes Refs. [39–54].

39. T. Matsunaga, N. Yamada, Y. Kubota, Structures of stable and metastable Ge$_2$Sb$_2$Te$_5$, an intermetallic compound in GeTe-Sb$_2$Te$_3$ pseudobinary systems, *Acta Crystallogr., Sect. B* **60**, 685 (2004).

40. S. Meng, E. Kaxiras, Electron and Hole Dynamics in Dye-Sensitized Solar Cells: Influencing Factors and Systematic Trends, *Nano Lett.* **10**, 1238 (2010).

41. N. Troullier, J. L. Martins, Efficient Pseudopotentials for Plane-Wave Calculations, *Phys. Rev. B* **43**, 1993 (1991).

42. J. P. Perdew, K. Burke, M. Ernzerhof, Generalized Gradient Approximation Made Simple, *Phys. Rev. Lett.* **77**, 3865 (1996).

43. J. L. Alonso, X. Andrade, P. Echenique, F. Falceto, D. Prada-Gracia, A. Rubio, Efficient Formalism for Large-Scale Ab Initio Molecular Dynamics based on Time-Dependent Density Functional Theory, *Phys. Rev. Lett.* **101**, 096403 (2008).

44. K. Sokolowski-Tinten, J. Bialkowski, and D. von der Linde, Ultrafast laser-induced order-disorder transitions in semiconductors, *Phys. Rev. B* **51**, 14186 (1995).

45. B. Lee, J. R. Abelson, S. G. Bishop, D. Kang, B. Cheong, and K. Kim, Investigation of the optical and electronic properties of phase change material in its amorphous, cubic, and hexagonal phases, *J. Appl. Phys.* **97**, 093509 (2005).

46. Q. Yin, L. Chen, Enhanced optical properties of Sn-doped $Ge_2Sb_2Te_5$ thin film with structural evolution, *J. Alloys Compd.* **770**, 692 (2019).

47. C. Steimer, V. Coulet, W. Welnic, H. Dieker, R. Detemple, C. Bichara, B. Beuneu, J. P. Gaspard, and M. Wuttig, Characteristic Ordering in Liquid Phase-Change Materials, *Adv. Mater.* **20**, 4535 (2008).

48. P. B. Moore, On the Relationship between Diffraction Patterns and Motions in Macromolecular Crystals, *Structure* **17**, 1307 (2009).

49. P. Fons et al., Picosecond strain dynamics in $Ge_2Sb_2Te_2$ monitored by time-resolved x-ray diffraction. *Phys. Rev. B* **90**, 094305 (2014).

50. A. V. Kolobov, P. Fons, J. Tominaga, and S. R. Ovshinsky, Vacancy-mediated three-center four-electron bonds in $GeTe$-$Sb_2Te_3$ phase-change memory alloys. *Phys. Rev. B* **87**, 165206 (2013).

51. Z. M. Sun, J. Zhou, Y. C. Pan, Z. Song, H. Mao, and R. Ahuja, Pressure-induced reversible amorphization and an amorphous–amorphous transition in $Ge_2Sb_2Te_5$ phase-change memory material, *Proc. Natl. Acad. Sci.* **108**, 10410 (2011).

52. X. P. Wang, X. B. Li, N. K. Chen, J. Bang, R. Nelson, C. Ertural, R. Dronskowski, H. B. Sun, and S. B. Zhang, Time-dependent density-functional theory molecular-dynamics study on amorphization of Sc-Sb-Te alloy under optical excitation, *NPJ Comput. Mater.* **6**, 31 (2020).

53. K. Y. Ding et al., Phase-change heterostructure enables ultralow noise and drift for memory operation, *Science* **366**, 210 (2019).

54. P. Zalden et al. Femtosecond x-ray diffraction reveals a liquid–liquid phase transition in phase-change materials, *Science* **364**, 1062 (2019).

55. L. Waldecker, R. Bertoni, R. Ernstorfer and J. Vorberger, Electron-Phonon Coupling and Energy Flow in a Simple Metal beyond the Two-Temperature Approximation, *Phys. Rev. X* **6**, 021003 (2016).

56. S. Sahu, A. Manivannan, and U. P. Deshpande, A systematic evolution of optical band gap and local ordering in $Ge_1Sb_2Te_4$ and $Ge_2Sb_2Te_5$ materials revealed by in situ optical spectroscopy, *J. Phys. D: Appl. Phys.* **51** 375104 (2018).

57. J. Bang, Y. Sun, X. Liu, F. Gao, and S. Zhang, Carrier-Multiplication-Induced Structural Change during Ultrafast Carrier Relaxation and Nonthermal Phase Transition in Semiconductors. *Phys. Rev. Lett.* **117,** 126402 (2016).

58. Luciana Vidas, Daniel Schick, Elías Martínez, Daniel Perez-Salinas, Alberto Ramos-Álvarez, Simon Cichy, Sergi Batlle-Porro, Allan S. Johnson, Kent A. Hallman, Richard F. Haglund, Jr., and Simon Wall, Does $VO_2$ Host a Transient Monoclinic Metallic Phase?, *Phys. Rev. X* **10,** 031047 (2020).

59. N. K. Chen et al., Giant lattice expansion by quantum stress and universal atomic forces in semiconductors under instant ultrafast laser excitation, *Phys. Chem. Chem. Phys.* **19**, 24735–24741 (2017).

60. P. Fons, H. Osawa, A. V. Kolobov, T. Fukaya, M. Suzuki, T. Uruga, N. Kawamura, H. Tanida, and J. Tominaga, Photoassisted amorphization of the phase-change memory alloy $Ge_2Sb_2Te_5$, *Phys. Rev. B* **82**, 041203(R) (2010).

61. K. V. Mitrofanov *et al.,* Sub-nanometre resolution of atomic motion during electronic excitation in phase-change materials. *Sci. Rep.* **6**, 20633 (2016).

62. Y. Qi, M. Guan, D. Zahn, T. Vasileiadis, H. Seiler, Y. W. Windsor, H. Zhao, S. Meng, R. Ernstorfer, Traversing double-well potential energy surfaces: photoinduced concurrent intralayer and interlayer structural transitions in XTe2 (X=Mo, W), *ACS Nano* **16**, 11124 (2022).

63. M. R. Otto, J.-H. Pöhls, L. P. René de Cotret, M. J. Stern, M. Sutton, B. J. Siwick, Mechanisms of electron-phonon coupling unraveled in momentum and time: The case of soft phonons in $TiSe_2$, *Sci. Adv.* **7**, eabf2810 (2021).

64. J. P. H. Rivett, L. Z. Tan, M. B. Price et al., Long-lived polarization memory in the electronic states of lead-halide perovskites from local structural dynamics. *Nat. Commun.* **9**, 3531 (2018).

65. C. Gehrmann, D. A. Egger, Dynamic shortening of disorder potentials in anharmonic halide perovskites. *Nat. Commun.* **10**, 3141 (2019).

66. E. S. Božin, C. D. Malliakas, P. Souvatzis, T. Proffen, N. A. Spaldin, M. G. Kanatzidis, and S. J. L. Billinge, Entropically Stabilized Local Dipole Formation in Lead Chalcogenides, *Science* **330,** 1660 (2010).

67. T. Keiber, F. Bridges, B. C. Sales, Lead Is Not Off Center in PbTe: The Importance of r - Space Phase Information in Extended X-Ray Absorption Fine Structure Spectroscopy. *Phys. Rev. Lett.* **111***,* 095504 (2013).

68. M. P. Jiang et al., The origin of incipient ferroelectricity in lead telluride, *Nat. Commun.* **7**, 12291 (2016).

69. B. Sangiorgio, E. S. Bozin, C. D. Malliakas, M. Fechner, A. Simonov, M. G. Kanatzidis, S.

J. L. Billinge, N. A. Spaldin, and T. Weber, Correlated local dipoles in PbTe, *Phys. Rev. Materials* **2**, 085402 (2018).

70. M. Paściak, T. R. Welberry, J. Kulda, S. Leoni, and J. Hlinka, Dynamic Displacement Disorder of Cubic BaTiO$_3$, *Phys. Rev. Lett* **120**, 167601 (2018).

71. K. Datta, I. Margaritescu, D. A. Keen, and B. Mihailova, Stochastic Polarization Instability in PbTiO$_3$, *Phys. Rev. Lett* **121**, 137602 (2018).

**ACKNOWLEDGEMENT**

We acknowledge Miquel Rudé and Valerio Pruneri at ICFO-Institut de Ciències Fotòniques for providing the GST-225 samples, Prof. Ming Xu and Meng Xu at Huazhong University of Science and Technology for helpful discussion. Yingpeng Qi acknowledges support by the Sino-German (CSC-DAAD) Postdoc Scholarship Program (Grant No. 201709920054 and No. 57343410). Hélène Seiler acknowledges support by the Swiss National Science Foundation under Grant No. P2SKP2 184100. Work in JLU was supported by the National Natural Science Foundation of China (Grants No. 61922035 and No. 11904118). High-Performance Computing Center (HPCC) at Jilin University for computational resources is also acknowledged. Yingpeng Qi and Ralph Ernstorfer initiated the project. Yingpeng Qi executed the experiments with the help from D. Zahn, T. Vasileiadis, and H. Seiler. Yingpeng Qi did the data analysis and the structure factor calculation. Nianke Chen and Xianbin Li did the TDDFT-MD simulation. Yingpeng Qi wrote the manuscript with contributions from all the authors.